\documentstyle[12pt]{article}
 \def\be{\begin{eqnarray}}
 
 \def\ee{\end{eqnarray}}
\renewcommand{\textfraction} {.01} 
\renewcommand{\topfraction} {.99} 
\input epsfig.sty
\begin{document} 
\pagestyle{empty}
\Huge{\noindent{Istituto\\Nazionale\\Fisica\\Nucleare}}

\vspace{-3.9cm}

\Large{\rightline{Sezione di ROMA}}
\normalsize{}
\rightline{Piazzale Aldo  Moro, 2}
\rightline{I-00185 Roma, Italy}

\vspace{0.65cm}

\rightline{INFN-1300/00}
\rightline{September 2000}

\vspace{1.cm}

\renewcommand{\textfraction} {.01} 
\renewcommand{\topfraction} {.99} 
 
\begin{center}{\large\bf  Neutron unpolarized structure function $F_2^n(x)$ from deep inelastic
scattering off $^{3}He$ and $^{3}H$ }
\end{center}

\begin{center} E. Pace$^a$, G. Salm\`{e}$^b$ and S. Scopetta$^c$\end{center}

\noindent { $^a$ \it Dipartimento di Fisica, Universit\`{a} di
Roma "Tor Vergata" and Istituto Nazionale di \\ Fisica Nucleare, Sezione Tor
Vergata, Via della Ricerca Scientifica 1, I-00133 Roma, \\Italy}

\noindent { $^b$ \it Istituto Nazionale di Fisica Nucleare, Sezione
di  Roma, P.le A. Moro 2, I-00185 Rome, Italy}

\noindent { $^c$ \it ECT*, Strada delle Tabarelle
286, I - 38050 Villazzano (Trento), Italy and Istituto\\  Nazionale di 
Fisica Nucleare, Sezione di Perugia, Via A. Pascoli, 00610 Perugia, Italy}

\vspace{1.5cm}

\begin{abstract}  
The possibility to safely extract the neutron deep inelastic structure function
$F_2^n(x)$ in the range $0 \le x \le 0.9$ from joint measurements of deep
inelastic structure functions of $^{3}He$ and $^{3}H$ is demonstrated. While 
the
nuclear structure effects are relevant, the model dependence in this extraction
linked to the $N-N$ interaction is shown to be weak.
\end{abstract}

\vspace{4.5cm}

\hrule width5cm
\vspace{.2cm}
\noindent{\normalsize{Proceedings of XVIIth European Conference on Few-Body
Problems in Physics, Evora (Portugal), Sept. 2000. To appear in 
{\bf Nucl. Phys. A}.}}

\newpage
\pagestyle{plain}

\vspace{1cm}
\indent 
The knowledge of both the proton and neutron deep inelastic structure functions
(DISF) at large x could give access to the valence
$u$ and $d$ quark distributions in the nucleon \cite{ud}. Usually deuteron 
data
have been employed to gain information on the neutron unpolarized DISF,
$F_2^n(x)$, but uncertainties remain.
In particular, it has been recently argued \cite{thom} that 
the standard treatment of deuteron 
data \cite{Bodek}, leading to the value $1/4$
for the ratio $r(x) = F_2^n(x)/F_2^p(x)$ when $x \rightarrow 1$,
could be unfair. An improved analysis could  
move such a value towards that of
$3/7$, suggested by pQCD arguments \cite{fj}, and closer to that of 
$2/3$, expected within SU(6) symmetry.

Recently, the possible use of an unpolarized $^3H$ target has been
discussed \cite{JCR}. In particular, an experiment has been proposed
\cite{MP}, aimed to the measurement of $F_2^n(x)$, using the ratio of
unpolarized structure functions of $^{3}He$, $F_2^H(x)$, and $^{3}H$,
$F_2^T(x)$, to reduce systematic errors in the measurements, as well as
theoretical model dependences.  

In this work a safe procedure is proposed to extract $F_2^n(x)$ from this
ratio \cite{PSS}, although it will be shown that, at high values of Bjorken
variable $x = Q^2 / (2 M \nu)$, nuclear structure effects are relevant and
cannot be overlooked. Nucleon structure in $^{3}He$ and $^{3}H$ is assumed to
be the same as for free nucleons, while nucleon motion and nucleon binding in
the three-nucleon systems are accurately taken care of {\sl and the Coulomb
interaction (CI) 
is explicitly considered in the evaluation of the $^{3}He$ wave
function}. For an easy presentation only the case of infinite momentum transfer
in the Bjorken limit will be considered, but it is straightforward to
generalize our approach to finite momentum transfer values \cite{PSS}.

 
The DISF for $^{3}He$ and for $^{3}H$ in impulse approximation (IA)
can be written 
as follows
\begin{eqnarray}
F_2^{H}(x) = 
2 \int\nolimits_{x}^{A} F_2^{p}(x/z) f_{p}^{H}(z) dz + 
\int\nolimits_{x}^{A} F_2^{n}(x/z) f_{n}^{H}(z) dz 
\label{34}
\end{eqnarray}
\begin{eqnarray}  
F_2^{T}(x) = 
2 \int\nolimits_{x}^{A} F_2^{n}(x/z) f_{n}^{T}(z) dz + 
\int\nolimits_{x}^{A} F_2^{p}(x/z) f_{p}^{T}(z) dz
\label{35}
\end{eqnarray}
where the distribution $f_{p(n)}^{H(T)}(z)$ describes the 
structure of the trinucleon system 
\cite{CLSPS}
\begin{eqnarray}
f_{p(n)}^{H(T)}(z) = 
\int\nolimits d E 
\int\nolimits d {\vec p} \; P^{H(T)}_{p(n)}( |{\vec p} |, E) \; \delta \left( z -
\frac {pq} {M\nu} \right) \; z \; C \; . 
\label{1}
\end{eqnarray}

\begin{figure}[t]
\vspace{6.2cm}
\includegraphics{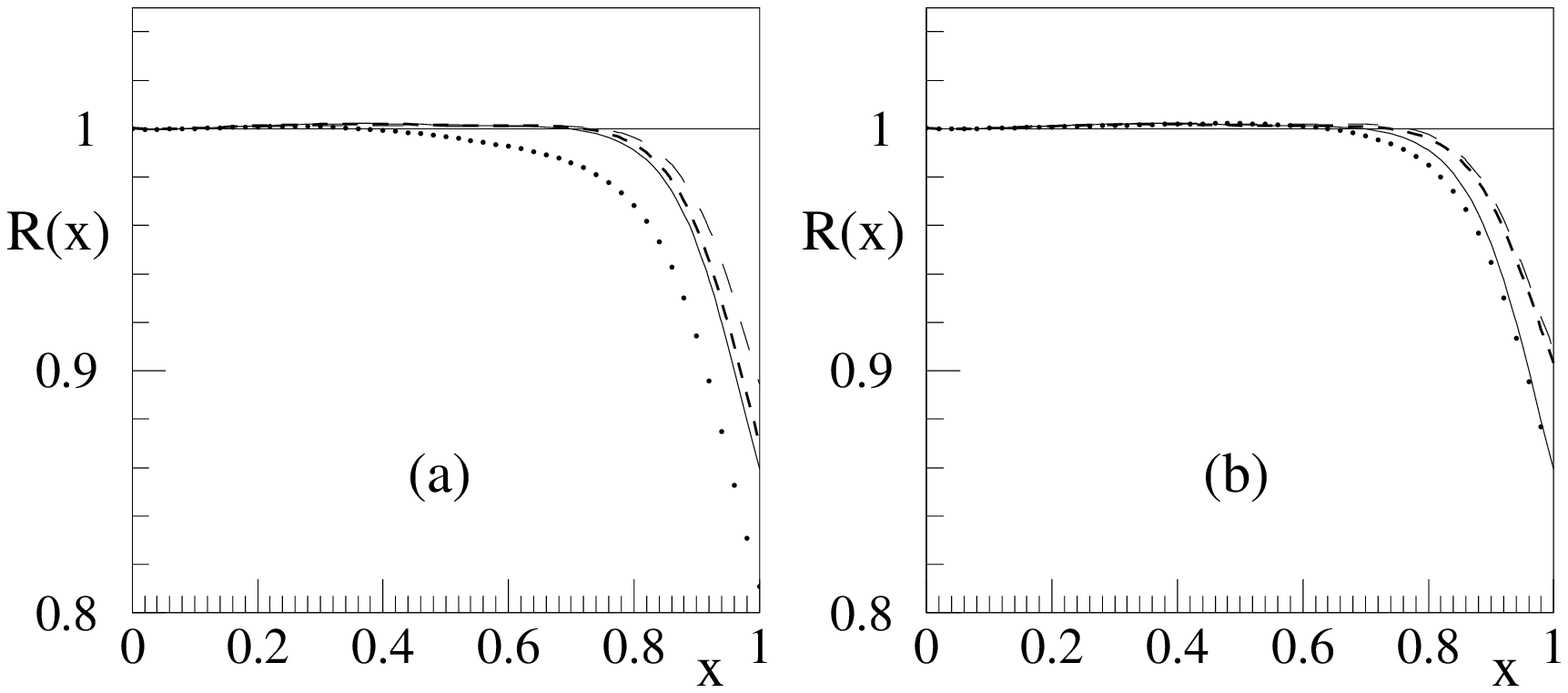}
\vspace{5mm} 
Figure 1. { (a) The super-ratio $R(x)$ (Eq. (\ref{37})) with $F_2^{n(p)}(x)$
from Ref. \cite{Aubert} for different $N-N$ interactions. Solid, dashed and
long-dashed lines correspond to the $RSC$ and $Av14$ interactions 
plus $CI$ and to $Av14$
without $CI$ for $^{3}He$, respectively. The dotted line is 
obtained as the solid one, but using the nucleon momentum distributions, 
instead
of the nucleon SF. (b) The super-ratio $R(x)$ for the $RSC$
interaction plus $CI$. Solid and dotted lines correspond to
the models of Ref. \cite{Aubert} and Ref. \cite{DL}
for $F_2^{n(p)}(x)$, respectively; dashed line: as the solid one with
$F_2^{n}(x)$ multiplied by a factor $(1 + 0.5 x^2)$; long-dashed line: $R(x)$
obtained with the arbitrarily modified $F_2^{n}(x)$, but evaluating the
corresponding variations of $R(x)$ only at the first order in $\delta \,
r^{(0)}(x) = 0.5 \, x^2 \; F_2^{n}(x)/F_2^{p}(x)$ (see text). }

\end{figure}
 
In Eq. (\ref{1}) the functions $P^{H(T)}_{p}(|{\vec p}|, E)$ and
$P^{H(T)}_{n}(|{\vec p}|, E)$ are the proton and neutron spectral functions
(SF) in $^{3}He$ ($^{3}H)$, respectively, ${\vec p}$ and $E$ the nucleon
momentum and removal energy, $M$ is the nucleon mass, and $C$ a
normalization factor. 

Let us define by $E(x)$ the experimental ratio between the $^{3}He$ and
$^{3}H$ unpolarized DISF :
$E(x) = F_2^H(x)/F_2^T(x) \;$,
and by $R(x)$ the super-ratio (S-R)
\begin{eqnarray}
R(x) = \frac{F_2^H(x)/(2F_2^p(x) 
+ F_2^n(x))}{F_2^T(x)/(2F_2^n(x) + F_2^p(x))} =
E(x) \; \; \frac{2 r(x) + 1} {2 + r(x)} \; \;. 
\label{37}
\end{eqnarray} 
In IA
the S-R is a functional of $r(x)$ ( $R(x) = R[r(x)]$ ).
Indeed from Eqs. (\ref{34},\ref{35},\ref{37}) one has
\begin{eqnarray}
R[r(x)] = \frac{2 \int\nolimits_{x}^{A} p(x,z) f_{p}^{H}(z) dz + 
\int\nolimits_{x}^{A} r(x/z) p(x,z) f_{n}^{H}(z) dz}
{\int\nolimits_{x}^{A} p(x,z) f_{p}^{T}(z) dz + 2 \int\nolimits_{x}^{A}
r(x/z) p(x,z) f_{n}^{T}(z) dz  } \; \;
\frac{2 r(x) + 1} {2 + r(x)}
\label{38}
\end{eqnarray}
where $p(x,z) = F_2^p(x/z) / F_2^p(x)$. From Eq. (\ref{37})
one can immediately obtain the following equation for the ratio $r(x)$ : 
\begin{eqnarray}
r(x) = \frac{E(x) - 2 R[r(x)]}{R[r(x)] - 2E(x)} 
\label{36}
\end{eqnarray}

The above equation is a self-consistent equation, which allows one to
determine $r(x)$.
If the distribution $f_{p(n)}^{H(T)}(z)$ is similar to a
Dirac $\delta$ function, $f_{p(n)}^{H(T)}(z) \sim \delta (z-1)$, then $R(x)
= 1$ and Eq. (\ref{36}) becomes trivial. Actually this hypothesis is only an
approximate one, so that $R(x) \neq 1$ at high $x$, as shown in Fig. 1.
Furthermore, sensible differences are obtained for $x \ge 0.6$, if the nucleon
SF are replaced by the nucleon momentum distributions. 
On the contrary, the model
dependence of $R(x)$, due to the $N-N$
interaction, is weak for any $x$ (see Fig. 1(a)). In particular,
the introduction
of a three-body force yields negligible
effects in
$R(x)$ at $x \le 0.90$.
\begin{figure}[t]
\vspace{6.2cm}
\includegraphics{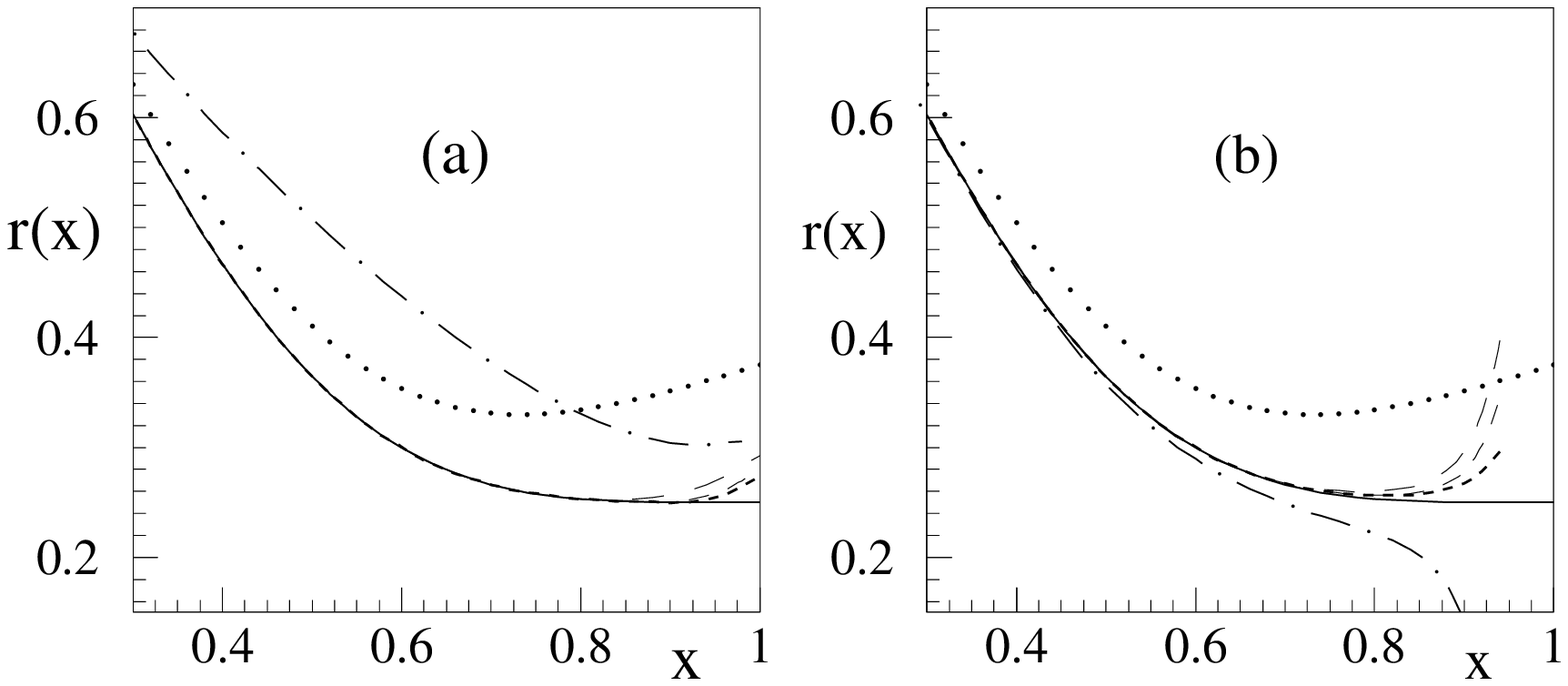}
\vspace{5mm}
Figure 2. { (a) The ratio $r(x)$ obtained from the recurrence relation
(\ref{39}), using the $RSC$ interaction plus $CI$ both for $E(x)$ and $R(x)$.
Solid and dot-dashed lines correspond to the nucleon DISF of 
Ref. \cite{Aubert} and \cite{DL}, respectively, while dotted
line corresponds to the model of Ref. \cite{Aubert}, 
multiplied by $(1 + 0.5 x^2)$ used as our $r^0(x)$ (see text). 
Long-dashed,
dashed, and short-dashed lines are $r^{(n)}(x)$ for $n = 3, 10, 20$,
iterations, respectively. 
(b) The same as in (a), but dashed, short-dashed, and long-dashed
lines are $r^{(20)}(x)$, obtained using for $R(x)$ the
nucleon SF obtained from the $RSC$ interaction without 
$CI$ for $^{3}He$, the $Av14$ interaction and the $Av14$
interaction without $CI$ for $^{3}He$, respectively
\cite{KPSV}. The dot-dashed line is the result of the iterative procedure
obtained using the nucleon momentum distribution for the evaluation of
$R(x)$. }     

\end{figure} 

 It is interesting to note that, if in Eq. (\ref{38}) $r(x)$ is replaced by
$r^{(0)}(x) = r(x) + \delta r^{(0)}(x)$, then the variations of $R(x)$ can be
accurately estimated using the expression obtained at the first order in
$\delta r^{(0)}(x)$ (see Fig. 1(b)). Thus, if in the right hand side of Eq.
(\ref{36}) the exact $r(x)$ is replaced by some approximation $r^{(0)}(x)$, the
expression for the super-ratio $R(x)$ at the first order in $\delta r^{(0)}(x)$
can be safely used. From this expression one easily realizes that the
values of $r(x)$ calculated from Eq. (\ref{36}) are closer to the exact ones
than the approximation $r^{(0)}(x)$. This observation allows one to solve Eq.
(\ref{36}) by recurrence    
\begin{eqnarray}  
r^{(n+1)}(x) = \frac{E(x) - 2 \, R[r^{(n)}(x)]}{R[r^{(n)}(x)] - 2 \, E(x)}  
\label{39}
\end{eqnarray}
starting from a reasonable zero order approximation for $F_2^n(x)$. 

To study the convergence of this recurrence relation, we evaluated both $E(x)$
and $R(x)$ using the nucleon SF obtained from the $RSC$ plus $CI$
interaction in Ref.
\cite{KPSV}. The nucleon structure functions of Ref. \cite{Aubert} were used in
$E(x)$, while for $R(x)$ the neutron one was arbitrarily modified by the factor
$(1 + 0.5x^2)$ to generate the zero-order approximation $r^{(0)}(x)$.
As shown in Fig. 2(a), a
sequence which rapidly converges to $r(x)$ of Ref. \cite{Aubert} is obtained
in the range $0 \le x \le 0.9$. In particular we find that up to $x = 0.9$
an accuracy better than $0.1 \%$ is obtained with
only ten iterations.
A convergence of a similar quality to the same function is also obtained using
for $r^{(0)}(x)$ the DISF of Ref. \cite{DL}.
On the contrary, if the momentum distribution is used for the evaluation of 
$R(x)$, instead of the SF, the iterative procedure converges to a
function $r(x)$, which differs from the correct one more than $10
\%$ for $x \ge 0.8$.

In order to check the model dependence of our approach, we repeated the whole
procedure, using for $R(x)$ SF corresponding to different
interactions than the $RSC$ one plus $CI$
used for $E(x)$. In the range $0 \le x \le
0.85$ the ratio $r(x)$ extracted by the recurrence relation after twenty
iterations differs from the one used for $E(x)$ less than $2.5 \%$, for any of
the considered $N-N$ interactions (see Fig. 2 (b)).
We have also succesfully applied our recurrence procedure to the extraction
of $r(x)$ from 
In conclusion the proposed procedure,
which has been successfully applied also to the $^2H-proton$ system
and to the $^3He-^2H$ system,
is able to yield reliable information on 
$F_2^n(x)$ in the $x$ range accessible at 
the upgraded TJNAF, whenever nucleon binding
in nuclei is correctly taken into account.

\end{document}